\documentclass[author,]{nrc1}
\usepackage{graphicx}
\begin{document}
\title{A historical perspective on Modified Newtonian Dynamics}
\author{R.H. Sanders}
\address{Kapteyn Astronomical Institute, University of Groningen}
\date{25 February 2014}
\shortauthor{R.H. Sanders}
\maketitle
\begin{abstract}
I review the history and development of Modified Newtonian Dynamics (MOND)
beginning with the phenomenological basis as it existed in the 
early 1980s.  I consider Milgrom's papers of 1983 introducing the idea
and its consequences for galaxies and galaxy groups, as well as the
initial reactions, both negative and positive.  The early
criticisms were primarily on matters of principle, such as the
absence of conservation laws and perceived cosmological
problems; an important step
in addressing these issues was the development of the
Lagrangian-based non-relativistic 
theory of Bekenstein and Milgrom.  This theory
led to the development of a tentative relativistic theory
that formed the basis for later multi-field theories
of gravity.
On an empirical level the predictive success of the idea with respect
to the phenomenology of galaxies 
presents considerable challenges for cold dark matter.  
For MOND the essential challenge remains
the absence of a generally accepted theoretical underpinning of
the idea and, thus, cosmological predictions.  I briefly
review recent progress in this direction.
Finally I discuss the role and sociology of unconventional
ideas in astronomy in the presence of a strongly entrenched standard
paradigm.  
\end{abstract}

\section{Introduction}

In the context of the present cosmological paradigm, $\Lambda$CDM,
there are two major
constituents of the Universe for which the only evidence
is astronomical.  There is dark energy, perhaps represented by a 
cosmological constant in Einstein's equations.  This medium,
comprising 70\% of the energy density of the Universe, causes the
observed present accelerated expansion evidenced by supernovae in
distant galaxies and makes up the energy difference necessary for
closure of the Universe.  And then there is cold dark
matter --  hypothetical particles, beyond the standard model of 
particle physics, comprising 25\% of the
Universe and interacting with baryonic matter (the remaining
5\%) primarily
through the force of gravity.  Because it is cold (i.e.,
non-relativistic at the time it decouples from the photons
and other relativistic particles in the early universe)
this medium promotes the early formation of structure on all
scales via
gravitational collapse and explains the discrepancy between
the directly observed baryonic matter (stars and gas) and the traditional
dynamical mass of bound gravitational systems such as galaxies
and clusters of galaxies.  Here the cosmological
paradigm impinges upon the dynamics of these well-observed local
systems and should, in principle,
be testable.  And here it fails.

The evidence supporting the standard cosmological 
paradigm is said to be so overwhelming that 
there is little room for doubt.  This is in spite of the fact
that
the most well-motivated dark matter particles -- supersymmetric
partners -- should be detectable in terrestrial  
experiments via the rare scattering of atomic nuclei.
In fact such events have never been seen in spite of considerable 
effort and expense invested in particle dark matter search experiments.
But efforts continue because 
$\Lambda$CDM has become something of an official religion -- 
a doctrine outside of which there is no salvation, beyond which
there is only damnation.
Yet there is a leading heresy that has attracted a relatively
small but growing number of adherents: modified Newtonian dynamics,
MOND, which, as will be argued here, is epistemologically ``more
correct" than CDM.

Viewed simply, MOND is an algorithm that, with one additional
fundamental parameter having units of acceleration, allows calculation
of the distribution of the effective gravitational
force in astronomical objects from the observed distribution of
baryonic matter -- and it works remarkably well.  
This is evidenced primarily by use of the MOND 
algorithm in the determination of rotation curves of disk  
galaxies where the agreement with observed rotation curves
is often precise, even in details.  The existence
of such an algorithm is
problematic for CDM because this is not something that
dissipationless dark matter on the scale
of galaxies can naturally do;
it would seem require a coupling between dark matter and baryonic
matter which is totally at odds with the perceived properties of CDM.

Moreover, MOND subsumes the Tully-Fisher law, the observed
near-perfect correlation between the baryonic mass of galaxies and
the asymptotic constant rotation velocity, as a aspect of
fundamental physics -- a Kepler's law for
galaxies.  This correlation appears unnatural in the context
of dark matter because the asymptotic rotation velocity
is a property of the dark matter halo that extends far beyond
the relatively puny concentration of baryons in the center.  
In the context of dark matter it is explained as resulting from
aspects of galaxy formation.  But it remains difficult to
understand how such a precise correlation can 
emerge from a process that must be inherently
quite random, with each galaxy having its own unique
history of formation, merging, feedback and dynamical evolution.

The success of the MOND algorithm has deeper implications
as a modification either of gravity (general relativity) or
of the way in which particles respond to an applied force at low
accelerations (inertia).  The idea will not be
generally accepted until these implications are understood
in the context of a more general theory, but the phenomenological
success should, in itself, be sufficient to raise serious
doubts about existence of cold dark matter and thus 
the prevailing cosmological paradigm.

Here my purpose to discuss MOND from a historical perspective, 
beginning with the phenomenological roots and philosophical
basis as outlined in the original papers of Milgrom 
some thirty years ago \cite{milg83a,milg83b,milg83c}.
I will describe the predictive successes of the idea and the 
initial and ongoing criticisms.  I will outline the
suggested physical bases of MOND while emphasizing that there
is not yet a generally accepted physical theory and, therefore,
cosmology.  I will be brief because many of these points
will be discussed in detail in the various articles in this
compendium, but I will conclude with a discussion of
the criteria that distinguish crazy from sensible ideas
in astronomy and of
the dangers to the creative process in science presented
by the unquestioning acceptance of a standard dogma. 

\section{The phenomenological roots of MOND}

The first rotation curves of spiral galaxies
measured in neutral hydrogen and extending well-beyond the 
visible disk were published by the mid 1970s \cite{shrog,robwhite}.  However,
to establish the fact that these extended
rotation curves did not decline in the expected Keplerian fashion
as well as the perception of this fact as a 
serious anomaly took 10 years more.  An important development in this 
realization was the demonstration by Kalnajs \cite{kaln} 
that for several galaxies with rotation curves measured in
optical emission lines, the shapes of the curves
were well matched by assuming that
the mass was distributed like the visible light from stars -- in the
context of Newtonian gravity.
This was confirmed by Kent \cite{kent1} for a sample of 37 bright
galaxies observed in optical emission lines by Rubin et al. 
\cite{rubetal}.  The
match, however, between observed and calculated rotation curves
did not extend beyond the visible disk where the gas kinematics
were measured only in the 21 cm line emission from
tenuous neutral hydrogen \cite{bosma,kent2}.  
Here, there
was a serious discrepancy between the traditional dynamical
mass and the directly observable mass in stars and gas. 
By 1985 the existence of this discrepancy 
had became irrefutable \cite{valbetal} (see my book \cite{dmp} on the
dark matter problem for a general discussion of these developments).  

The initial
reaction of galactic dynamicists was to ascribe this phenomenon to 
unseen or dark matter in a surrounding spheroidal halo, a construct
already in place as a means of taming the instability of
rotationally supported disk systems \cite{op}. 
The tendency of
rotation curves to be asymptotically flat suggested
that the dark matter halo in galaxies had the density law of an 
isothermal spheroid, i.e., density falling like $1/r^2$.  
Perhaps this dark matter was of the same sort that contributed to
the virial discrepancy in clusters of galaxies -- a discrepancy 
first identified by Zwicky \cite{zwick} four decades earlier.  

The nature
of the dark matter was an issue for speculation (as it still is),
but initially low luminosity baryonic
objects were preferred -- low mass stars, dead stars 
(white dwarfs or neutron stars), planetary mass objects, 
low mass black holes,
snow balls.  But increasingly, it became appreciated that 
some sort of universal non-baryonic particle matter 
was needed to promote
the formation of the observed structure in an expanding
Universe of finite lifetime --
massive neutrinos, supersymmetric partners 
(weakly-interacting-massive particles, WIMPS), or axions
became the preferred choice for the cosmological as well
as the cluster and galactic
dark matter. 

At the same time that the existence of a discrepancy in
galaxies was beginning to be appreciated, certain regularities in
galactic photometry and kinematics were being revealed.
The most important of these was the Tully-Fisher relation,
a correlation between the luminosity of spiral galaxies and
the width of the global 21 cm line emission of individual
galaxies, a measure of the rotation velocity \cite{tf}.  
The form of the
correlation is a power law, $L \propto (\Delta V)^\alpha$
where $L$ is the luminosity of the galaxy in a particular
photometric band and $\Delta V$ is the line width corrected
for inclination of the galaxy disk.  
The exponent $\alpha$ depended upon the color in which
$L$ was measured and ranged from 2 in the blue light (B-band) 
to 4 in the
near-infrared.  Because of its potential use as a distance
indicator, considerable effort was spent upon calibration
of the relationship, and, by 1980 it had been established
that the the near infrared H band ($\approx$ 1.6 $\mu$m) is
the preferred color for measuring the luminosity.  This
is near the peak of the black body radiation from old
stars making up the bulk of the stellar population and, unlike
the B-band emission, is 
relatively free of the effects 
recent star formation and dust absorption \cite{ahm}.  
The scatter about the mean relation is much reduced in the
near-infrared, and, because the near-infrared luminosity
is most nearly proportional to the stellar mass (near constant
mass-to-light ratio), this implies
that the true relation is between stellar mass and rotation
velocity.

Faber and Jackson \cite{fj} had discovered a similar relation for
elliptical galaxies, systems supported not so much by
systematic rotation but by random motion of the stars. Here the 
relation is between the luminosity and
the velocity dispersion of the stellar component
and has the form $L\propto \sigma^\beta$
where it is also the case that $\beta \approx 4$.  The observed
relationship has greater scatter, but is also an indication of
a more fundamental correlation between mass and velocity dispersion
-- a relation going beyond the Newtonian virial theorem with
three parameters.

A regularity in the photometric properties
of spiral galaxies had been discovered even earlier.  In 1970
Freeman \cite{kfree} noted that the
distribution of surface brightness in the disks of
spiral galaxies can be generally described as having
an exponential form and that the mean surface 
surface brightness
within an exponential scale length appears to have a characteristic 
value of about 21.6 magnitudes
per square arc second in the B band corresponding to
roughly 150 solar luminosities per square parsec.  Later it
was argued \cite{alshu} that this was actually
an upper limit and not an average value; that is to say, there
are fainter (lower surface brightness) galaxies but not 
brighter ones.  Again for elliptical galaxies there is
a similar characteristic central surface brightness 
averaged within an effective radius \cite{fish} -- that radius containing
half the luminosity.

Initial spectroscopic measurements of the radial velocities
in several stars in very
low surface brightness systems, such as the dwarf spheroidal
companions of the Milky Way, were carried
out in the early 1980s.  There were early indications, but only hints
by 1983 \cite{dwfsph}, that the dynamical mass to light ratio in such
systems might be large.  By 1982
several low surface brightness spirals had been discovered 
\cite{rometal};
initial 21 cm line observations indicated that these objects
might have a somewhat larger mass-to-light ratio
than high surface brightness galaxies, but there was no
suggestion of 
a surface brightness dependence of the dark matter
content of galactic systems.

This is where the observational situation stood around 1982.
In flat rotation curves, extending well beyond the visible
disk, there was emerging evidence for a discrepancy between the 
visible and classical dynamical mass -- a discrepancy that
grew in the outer regions and, if
described by dark matter, required a $1/r^2$ density
relation in a dark halo
(or a $1/r$ surface density if described by a dark disk).
There were, in addition, scaling laws and photometric
regularities of galactic
systems:  the Tully-Fisher law for spirals and Faber-Jackson
for ellipticals; a characteristic surface brightness
for spirals and for ellipticals, but there was
no clear indication of a
surface brightness dependence to the
discrepancy in galaxies.  This set the stage for the
appearance of MOND as an alternative to dark matter,
but before describing this development, it is of 
interest to discuss a diversion along this path.

\section{A modification at large length scale}

The discrepancy is apparent in large astronomical systems --
the outer regions of spiral galaxies and clusters
of galaxies.  On the scale of the solar system, there is
no evidence at all for any deviation from Newton's law.
Therefore, an obvious modification is to change
the law of attraction beyond a length scale, $r_0$, of
galactic dimensions. 
This would appear to be consistent with a larger discrepancy in
larger systems -- clusters of galaxies, for example,
would have more missing mass than individual galaxies, which seemed
to be the case before the discovery of hot gas in clusters.

In 1963 Finzi \cite{finzi} emphasized the 
ubiquity of the mass discrepancy
in large astronomical systems.  He pointed out that in the Milky
Way the total mass appeared to grow with scale (this from work
on the kinematics of globular clusters \cite{kurth}).  Early 21 cm line 
observations of the rotation curve of the Andromeda galaxy,
indicated the same \cite{vdhetal}.  In the largest bound gravitational systems,
the great clusters of galaxies, the mass-to-light ratio 
appeared to approach several hundred.  Finzi first 
stressed the apparent universality of these phenomena
and noted that these observations could be
explained with a modification of Newtonian attraction
which fell more as $1/r^{1.5}$ rather than $1/r^2$ beyond
a scale of about 1 kpc.

By the mid-1980s it had became generally apparent that rotation curves
of spiral galaxies were asymptotically flat.
Flat rotation curves will result if the gravitational
acceleration about a point mass varies as
$$f = \nu(r/r_0)GM/r^2 \eqno(1)$$
where  
$\nu(x)$ is a function with the asymptotic behavior such that 
$\nu(x) \rightarrow 1$ when $x<<1$ and $\nu(x) = x$
when $x>>1$.  Then in the limit of large distance
($r>r_0$), $f=GM/(rr_0)$.  Equating this to the
acceleration of circular motion yields a constant
rotation velocity $$V^2 = GM/r_0. \eqno(2)$$

Such modifications were discussed by Tohline \cite{tohline} and
Kuhn and Kruglyac \cite{kk87}.  I proposed a specific
example in which the gravity force was the sum of
two components, an attractive force of infinite extent
and a repulsive component mediated by a particle of
finite mass and hence finite extent \cite{rhs84}.  
In this case he
gravitational potential would be
$$U(r) = {{GM}\over{r}}(1-\alpha e^{-r/r_0}). \eqno(3)$$
If $\alpha \approx 0.9$ then this would yield
a plateau of constant rotation velocity from 0.5$r_0$ to
2.5$r_0$ followed by a return to $1/r^2$ attraction and
a Keplerian decline on larger scale.  I felt that
$1/r^2$ attraction on largest scale would lead to a 
more sensible cosmology because it would permit
a maximum discrepancy as well as the usual
Newtonian derivation of the Friedmann equation.

The essential problem, which should have been already obvious
at this point, is that modifications of this sort are in 
direct contradiction to the observed 
Tully-Fisher relation.  Eq. 2 implies a
relation of the form $M\propto V^2$ whereas the exponent in the
luminosity-rotation velocity relation observed at that
time was certainly greater than 
three.  One might suggest that the mass-to-light ratio
varies systematically with galaxy mass, i.e., $M/L\propto V^{-2}$; 
that is to
say, more massive galaxies have a smaller M/L, but 
the tendency is just the opposite.  It appears that,
if anything, more massive galaxies have a larger M/L
because these tend to be earlier type bulge-dominated
systems.  Moreover, in the near-infrared where
M/L is relatively constant,
the exponent is most nearly 4 and the scatter is 
lowest.  

In addition, any such 
modification would imply that smaller galaxies should
have a smaller discrepancy and larger galaxies a 
larger discrepancy.  Even by 1980, it had become
obvious that this was not the case.  In a 1979 review Faber
and Gallagher \cite{fg79} tabulated the dynamical masses and mass-to-light
ratios of 
50 nearby galaxies;  no correlation between size (as
measured by photometric radius) and M/L was noted.
This is evident in the left hand panel of Fig.\ 2 which
is based upon more recent data \cite{sanverh}.  This shows the dynamical
M/L determined via the full 21 cm line rotation curves plotted
against the
galaxy size (the radius at the last measured point of the
rotation curve) for a sample of galaxies in the Ursa Major
cluster.

Thus the early evidence did not support a modification of
gravity beyond a length scale, although this fact did not
prevent myself and others from proposing such  
hypotheses -- and this after the idea of MOND had been
published.

\section{An acceleration-based modification}

Modified Newtonian dynamics is solely the invention of
Mordehai (Moti) Milgrom.  The idea of an
acceleration-based modification of dynamics or 
gravity would have probably occurred to someone else
sooner or later, but it is safe to say that in the early 1980s
no one but Milgrom had considered such a possible modification
as an alternative to astrophysical dark matter.
It was a brilliant stroke of insight to realize that
astronomical systems were not only characterized by
large scale but also by low internal accelerations
and that this could account for the known systematics
in the kinematics and photometry of galactic systems.
However, the idea was hardly greeted with overwhelming 
enthusiasm. 

Before 1980 Milgrom (Weizmann Institute, Israel)
had worked primarily in high-energy
astrophysics and was well-known for his highly
successful kinematic model of SS 433, a compact object
with precessing  high velocity jets.  During a 
sabbatical year at the Princeton Institute for Advanced
Study, he decided to consider the dynamics of galaxies and
galaxy systems in general and
the dark matter problem in particular, taking advantage of 
the considerable expertise on this subject at Princeton.  
He felt that so long as the only evidence for dark
matter on astronomical scales 
was its putative global gravitational or dynamical 
effects, then its presumed existence is not independent of the
assumed law of gravity or dynamics on those sales --  
that so long as no candidate dark matter 
objects or particles had been identified, then it was
legitimate to look for alternative solutions to
the discrepancy in modifications of Newtonian gravity or dynamics.
Such a point-of-view seems hardly radical at all but
an entirely reasonable scientific approach.

Looking at the existent observations, Milgrom realized that
a distance-based modification could not work for those reasons
noted above.  He also realized that an acceleration-based
modification could account for these observations, in particular
the systematics of galaxy photometry and kinematics.
Thus he proposed a simple rule that can be realized
as a modification of the law of inertia:  the inertial reaction
of a particle of mass $m$ to an applied force $\bf F$ is 
$${{\bf F}\over{m}} = {\bf a}\mu(a/a_0) \eqno(4)$$ where $\bf a$ is the resulting
acceleration and $a_0$ is a new fixed 
parameter with units of acceleration.   The function $\mu(x)$
is not specified but must have the asymptotic behavior such
that $\mu(x) = 1$ in the limit where $x\gg 1$ (to retain 
Newtonian dynamics) but $\mu(x) = x$ where $x\ll 1$.
The rule may also be described as a modification of the Newtonian
gravitational acceleration ${\bf g_N}$; the true gravitational acceleration
${\bf g}$ is given by
$${\bf g}\mu(g/a_0) = {\bf g_N}.\eqno(5)$$

With respect to particles in orbital motion in a galaxy the
two formulations are equivalent but, obviously, there are 
differences in principle since ${\bf g}$ refers to specifically
the gravitational acceleration but ${\bf F}$ can be any applied
force.  The relationship between the applied force per unit
mass and the resulting acceleration is shown in Fig.\ 1; at
accelerations prevailing in the Solar System, Newton and MOND are indistinguishable.

\begin{figure}
\begin{center}
\includegraphics[height=8cm]{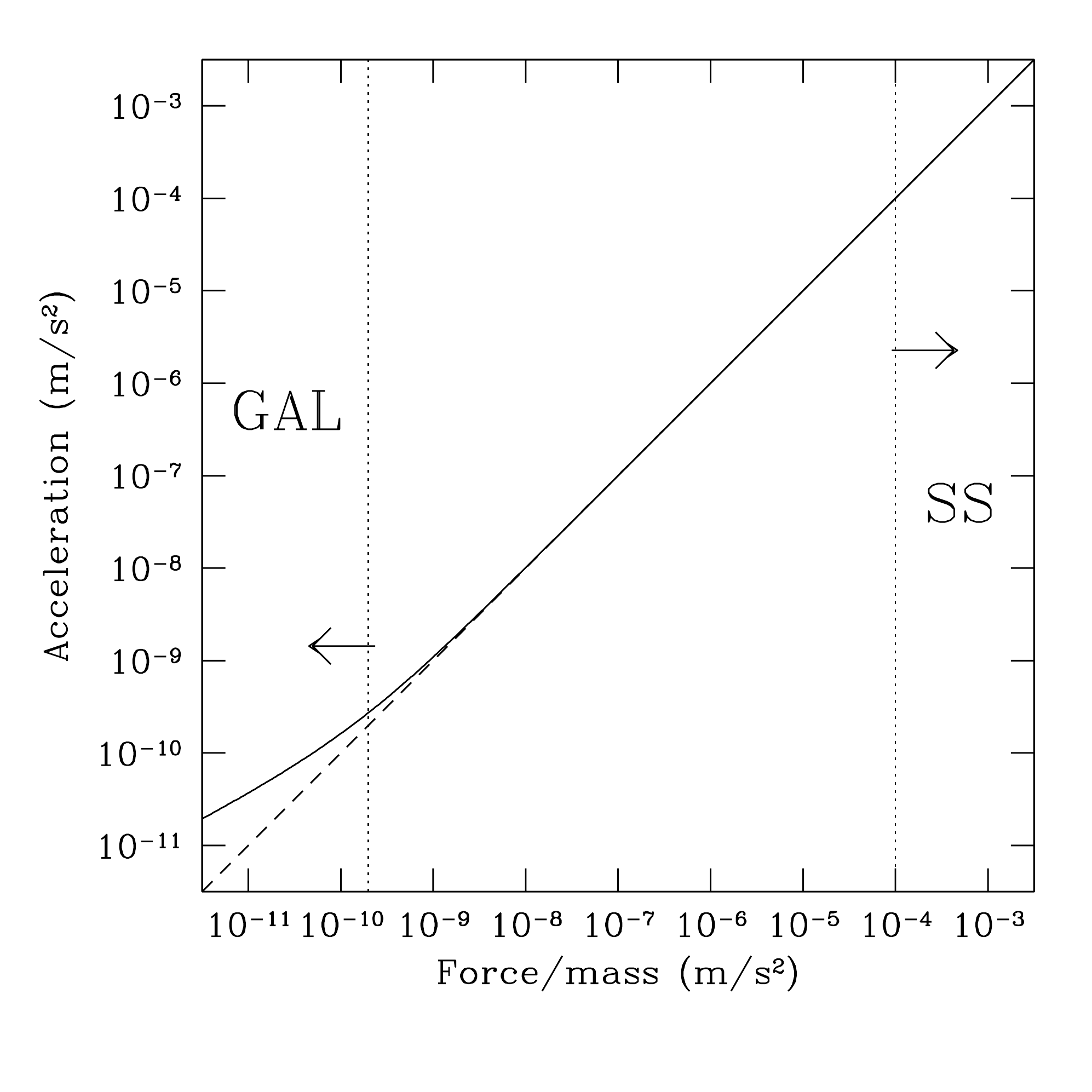}
\caption{On this logarithmic plot the acceleration resulting from an applied force
(per unit mass) is shown for Newtonian dynamics (dashed curve) and
for modified Newtonian dynamics (solid curve).  These only
differ significantly below an acceleration of $10^{-10}$
cm/$s^2$.  The indicated
region on the right corresponds to accelerations in the Solar System \index{Solar System}
and that on the left to accelerations in the outer regions of bright
galaxies.  Solar System \index{Solar System} accelerations are deep in the Newtonian
regime, but galaxy scale accelerations are typically 
in the modified regime.  Here it is assumed that $\mu(x) = x/(1+x)$.}
\end{center}
\end{figure}

The hypothesis is that the discrepancy should begin at an
acceleration below the critical acceleration $a_0$. 
There is very convincing evidence from 21 cm line rotation curves
observed in the late 1990s \cite{sanverh}
that this is the case.  In Fig.\ 2, right panel, we see the
same sample of galaxies from the Ursa Major cluster \cite{sanmcg} 
but this
time with the Newtonian dynamical mass-to-light plotted
against centrifugal acceleration ($V^2/r$) in units
of $10^{-10}$ m/s$^2$ at the last
measured point of the rotation curve.  Keeping in mind that the
stellar mass-to-light ratio (in solar units)
of spiral galaxies in the near-infrared
is on the oder of unity, we see that the
discrepancy is clearly correlated with acceleration in the
sense that the dynamical mass grows below accelerations
of one in these units, i.e., $a_0 \approx 10^{-10}$
m/s$^2$. It is important to recall that observations of
this quality did not exist when Milgrom made his proposal.

\begin{figure}
\begin{center}
\includegraphics[height=7cm]{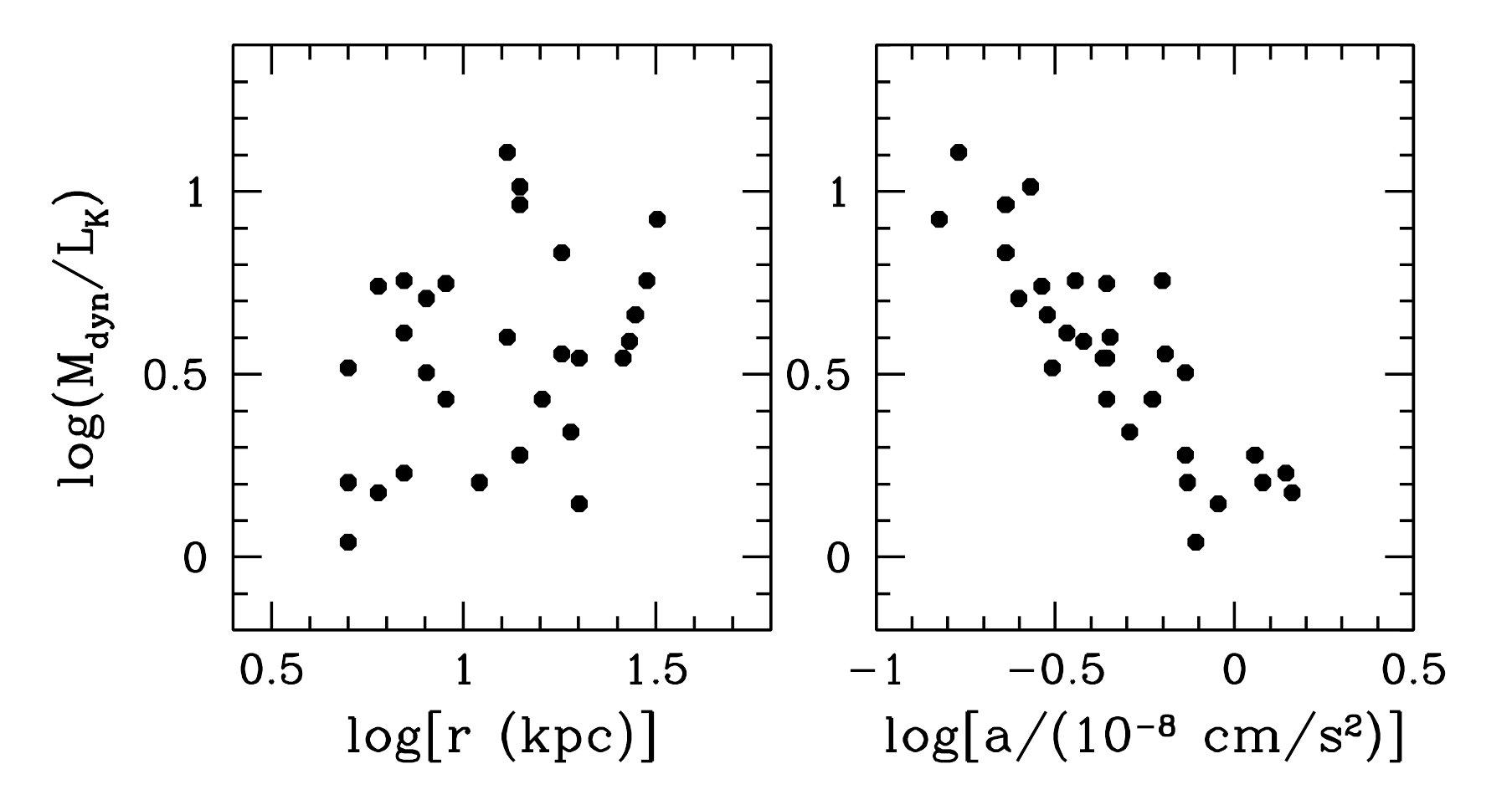}
\caption{The Newtonian mass-to-K'-band-luminosity
ratio of Ursa Major spirals at the last measured point
of the rotation curve plotted (on left) against the radial
extent of the rotation curve and then (right) against the centripetal
acceleration at that point \cite{sanverh,sanmcg}.}
\end{center}
\end{figure}

In the case of circular motion in a Newtonian force field 
in the low acceleration limit it follows directly from
eqs.\ 4 or 5 that 
$$V^4=GMa_0. \eqno(6)$$
In other words, the rotational velocity is constant and
related to the mass as $V^4 \propto M$.  Thus flat 
rotation curves and the observed Tully-Fisher law are
subsumed by the MOND algorithm.

\begin{figure}
\begin{center}
\includegraphics[height=7cm]{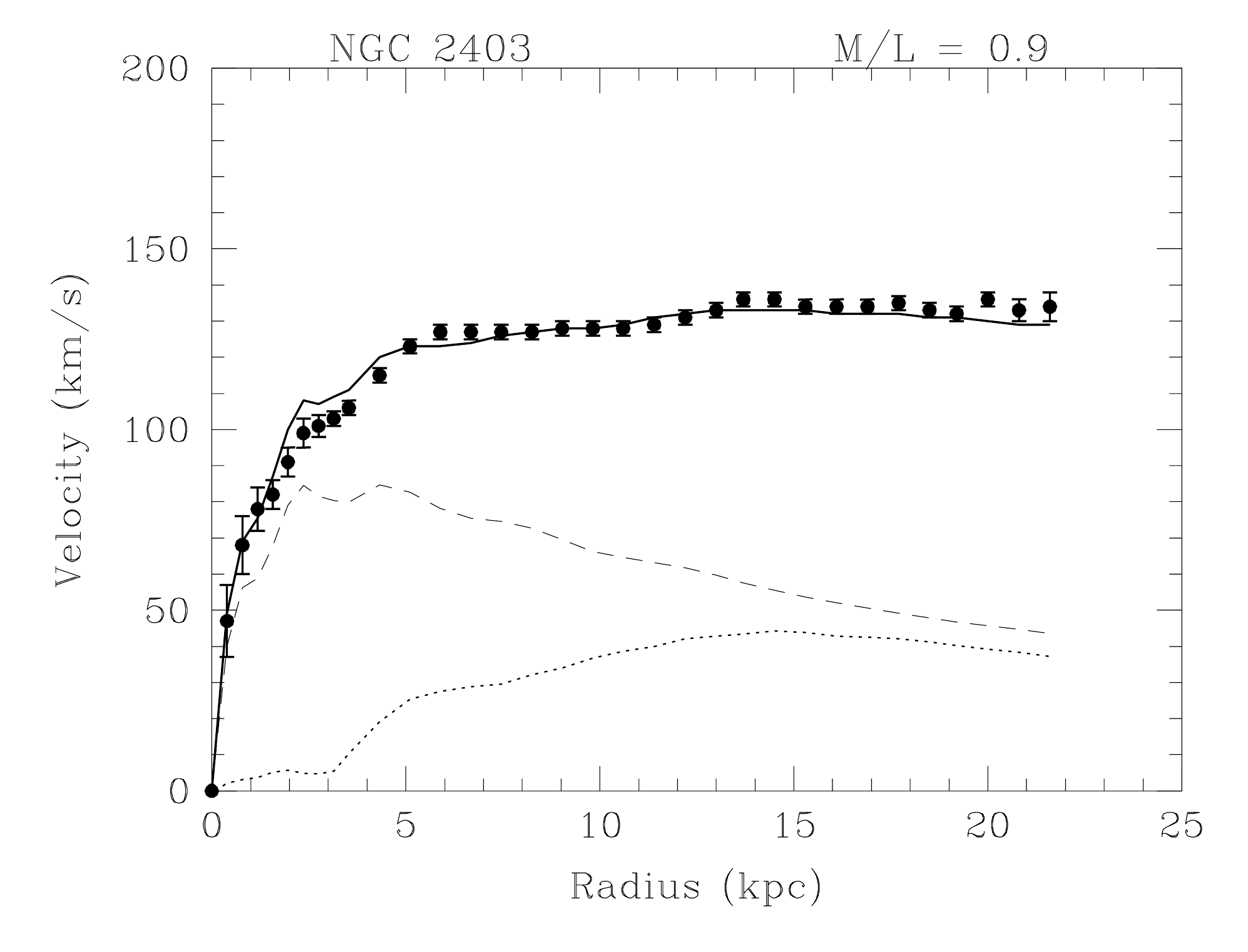}
\caption{The rotation curve of NGC 2403.  The points
are the observed rotation curve, the dashed and dotted
curves are the Newtonian rotation curves of the baryonic
components (stars and gas respectively), and the solid curve
is the MOND rotation curve \cite{bbs}.}
\end{center}
\end{figure}

These two aspects of MOND, flat rotation curves and a Tully-Fisher
relation, were part of the observed phenomenology before
MOND was written down, so it might asked if they 
can they rightly be called predictions.
I believe that it is fair to characterize them as
such because, in the context
of MOND, these attributes, following from physical law, 
are absolute:  every rotation curve of an isolated spiral galaxy must  
be {\it asymptotically} flat (the prediction of asymptotic flatness is 
important because by 1982 no one claimed this as a property
of rotation curves).  And every galaxy, without ambiguities
introduced by uncertainties
of inclination, distance or tidal disturbance, must lie on
{\it the} Tully-Fisher relation, {\it {so long as the asymptotic 
velocity is plotted against the detectable baryonic mass}}.
That is to say, Tully-Fisher is not a relation between
the maximum or an averaged rotation velocity and the luminosity
in some appropriate color band, but between the asymptotic rotation 
velocity and the baryonic mass of the galaxies \cite{mcgbtf}.  When these
are the plotted properties, there should be no intrinsic
scatter, and this, in fact, is a powerful prediction.

\begin{figure}
\begin{center}
\includegraphics[height=7cm]{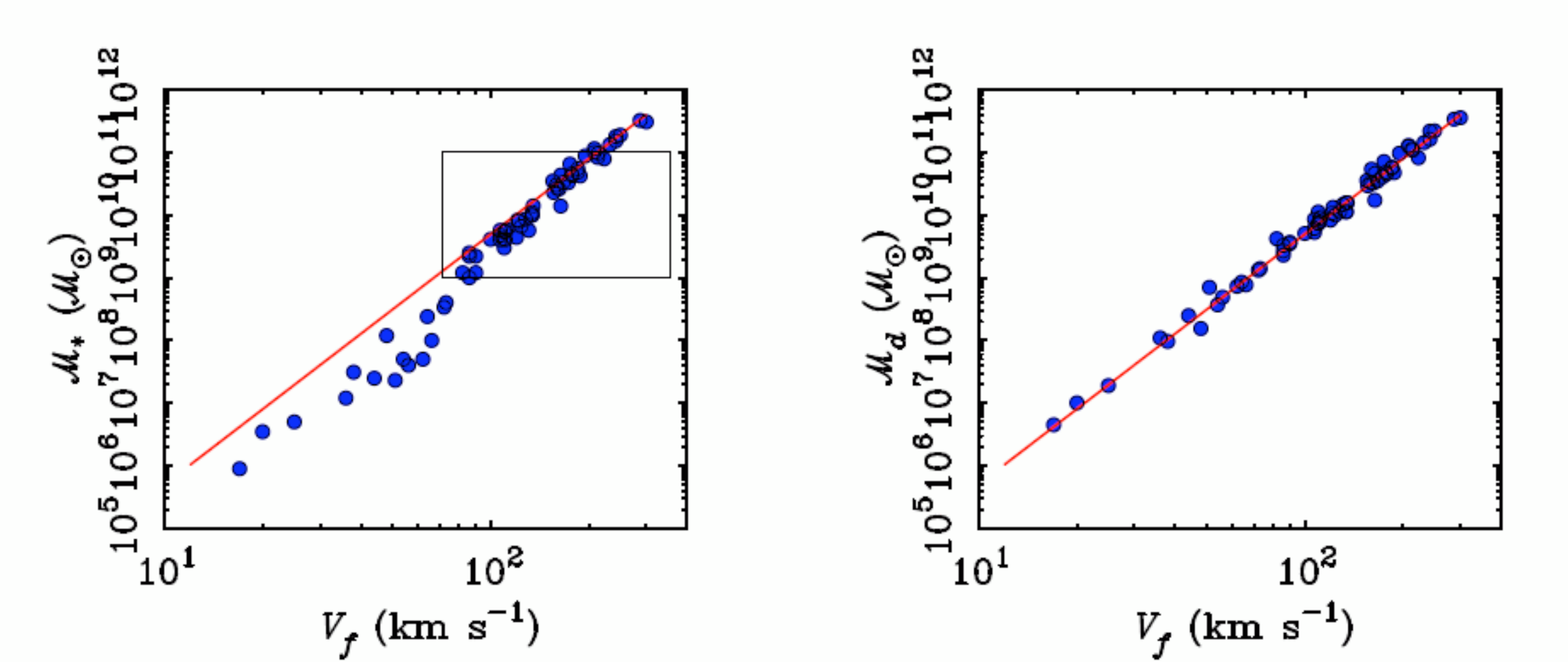}
\caption{The baryonic Tully-Fisher relation.  The panel
on the left is the mass in stars plotted against
the asymptotic rotation velocity of spiral galaxies.
On the right, the mass in gas is included \cite{mcgbtf}.}
\end{center}
\end{figure}

The results of modern observations are shown in Figs.\ 3 and
4.  Here the observed rotation curve (Fig.\ 3) demonstrates the
well known asymptotic behavior beyond the visible object.
The solid curve is the rotation velocity predicted by
MOND with $a_0 = 10^{-10}$ m/s$^2$, and the dashed and 
dotted curves are the Newtonian rotation curves resulting
from the stars and gas respectively.  The baryonic
Tully-Fisher relation is shown in Fig. 4, and its normalization,
given by $Ga_0$ again implies $a_0 = 10^{-10}$ m/s$^2$ \cite{mcgbtf}.
The interesting aspect of this plot is that in
the left-hand panel only the mass of the stars (estimated
from population synthesis models) is plotted against the
asymptotic rotation velocity.  The lower mass systems
are dominated by gas and they fall below the relation 
determined by the higher mass, star dominated systems.
When the gas mass is included, as in the right-hand panel,
the low mass systems lie on the same TF relation demonstrating
vividly that the true relation is between {\it baryonic}
mass and rotation velocity.

As Milgrom pointed out, 
the numerical value of the critical acceleration 
is $\approx cH_0/6$.  This is certainly not just coincidental
and suggests a connection of local dynamics with cosmology,
perhaps via a cosmological derivation of Mach's Principle.
It is also of interest that that $a_0$ is the maximum
gravitational acceleration on a subatomic scale; it 
would be the acceleration of gravity at one Compton radius
from a particle with a mass of several hundred MeV.
Combined with the proximity of $a_0$ to $cH_0$ yields
the famous Dirac cosmological numerical coincidences.

Apart from these aspects of galaxy phenomenology that follow
directly from the MOND hypothesis there are
additional observational properties that are clear predictions.   
The critical acceleration, $a_0$,
can be expressed as a surface density,
$\Sigma_0 = a_0/G$, which emerges as a characteristic
value.  With a mass-to-light ratio on the order of
unity, this can be translated into a critical surface
brightness that is comparable to that of Freeman 
and of Fish \cite{kfree,fish}.  
Given the well-known instability of rotationally
supported Newtonian disks \cite{op} this implies the critical 
surface brightness should appear as an upper limit
(disks with lower surface brightness are not Newtonian).  
Moreover, objects such as globular clusters
or elliptical galaxies with higher surface brightness,
should exhibit little discrepancy between the observed
and classical dynamical mass within the bright visible 
object.  On the other hand, low-surface brightness
objects are entirely in the low acceleration regime and
should exhibit a large discrepancy.  So there should be
a strong surface brightness dependence to the apparent dark matter
content of systems.  Now this is known to be the case, but
in 1982 it was a bold prediction.

There was an additional prediction concerning the general
shapes of rotation curves: in
high surface brightness spiral galaxies
that are near Newtonian in the bright inner regions,
the rotation curve should decline to the asymptotic
constant value beyond the bright disk;  in 
low surface brightness spirals however, the rotation curve
should slowly rise to the constant value in the
outer parts.  Again in 1983 
these systematics had never before been observed; this
property of rotation curves was pointed out explicitly in
1991 \cite{casvg}. 

Binary galaxies, small groups, and clusters of galaxies, should,
in so far as they are in the low acceleration regime,
also present large discrepancies.  It is now known that 
the central regions of rich clusters of galaxies are 
in the high acceleration regime, so here MOND predicts more
mass than is directly observed (the dynamical discrepancy is
not removed in the central regions), but in 1980 this was 
not evident (more on this issue below).

In 1983, Milgrom published three papers in the {\it Astrophysical
Journal} that fully
described the idea of MOND and these observational consequences [1-3].  
The first
was a short paper that introduced the idea;  the second dealt
with galaxy phenomenology and the third with groups and clusters of
galaxies.  Getting these papers past critical referees
was not an easy task.  Milgrom had previously submitted
the introductory paper to {\it Nature} and to {\it Astronomy and
Astrophysics Letters} without success.  The final three published
papers were received by the Astrophysical Journal in February of 1982 which
implies that the work had been done in 1981 or earlier -- truly
remarkable considering the tentative state of the data at that
point. 

Most of the 
criticisms were of matters of taste or principle rather than
phenomenological.  Typical is the objection of one
referee of the short paper originally submitted to
Astronomy and Astrophysics
Letters: ``In judging the success of any such proposed
modification one must draw up a balance sheet of gains
and losses.  In this theory there are very considerable
losses of accurately checked phenomena in order
to achieve an interpretation of phenomena that are not
well understood while maintaining that `most of what there
is, can be seen' and so dispensing with hidden
matter. " The referee goes on to say that to
obtain this dubious gain one loses various 
cherished principles such as equivalence, relativistic
invariance, etc.  ``My personal 
opinion is that such speculations are fun around the
coffee table, but I don't want to see them in
the literature until they are significantly further
developed."  In other words, before a more complete theory
can be presented, forget about publication.  

In his response to this argument, Milgrom mentioned
the Bohr atom.  ``It was not even a theory but a set
of assumptions on what an electron does.  It gave answers
to a very small number of questions, etc., etc.  It took
13 years and many good men to bring quantum mechanics to
the stage of the Schroedinger equation, which was not even
relativistic, which disposed of the notion 
that mechanics is deterministic and other such well 
established principles."  So, following Milgrom, if a
theory is successful in explaining diverse phenomena,
as is MOND (but actually not the Bohr atom!), then it should be
taken seriously even if it abandons cherished principles.
It would not be the first time that ``fundamental principles"
have been abandoned (or enlarged) in progressing to new physics.

But the referee makes an additional point:  the phenomena
that Milgrom's theory attempts to explain are not 
well-established, unlike the principles he abandons.
It is actually true that these systematics -- 
the universality of asymptotically
flat rotation curves, the Tully-Fisher relation and its preferred
exponent of four, the existence of a preferred surface brightness
and the presence of a large discrepancy in lower surface
brightness objects -- were not generally accepted or known or
appreciated when Milgrom wrote down his theory.
But this is where he showed a particular prescience --
the ability to glean from the data those systematics that 
are significant and that would be verified by coming 
observations, and the connection of these systematics through
a single preferred universal value of acceleration.

This referee and several others pointed out that
the original algorithm did not conserve the linear
or angular momentum of an isolated compound system.
To many, the idea of a binary star accelerating
away by itself, seemed like distinctly bad physics.
Milgrom was, from the beginning, aware of this
deficiency and, with Bekenstein, set about demonstrating
that it was not a necessary property of this
sort of theory.  The complaint, however, remained for 
several years (Felton \cite{felt} in 1984 was the first to point this 
out in print).  In the same vein, the motion of a compound
object in an external field was murky.  A star, for example,
in the outer galaxy is moving in a low acceleration 
external field.  Yet the internal motion of the gas
particles comprising the star feel an acceleration 
well in excess of $a_0$.  How does the star move?
Is it the acceleration of the center-of-mass that counts,
or is it the acceleration of the individual particles?

Not everyone was hostile to Milgrom's idea.  There were
several eminent scientists, who, if not actual supporters,
had some appreciation of his proposal, who thought
it was a legitimate avenue for research, and who thought that
the idea should be published.  Among these were Ed Salpeter
of Cornell and Scott Tremaine of Princeton.  But both
Tremaine and Salpeter pointed out a substantial phenomenological
problem:  galactic or open star clusters in the plane of
the Milky Way have accelerations below $a_0$ and
that should put them in the low acceleration limit.  But
these objects have no significant measurable discrepancy;  there is
no missing mass problem for galactic star clusters.
This criticism lead Milgrom to the realization of presence of the external
field effect; that is to say, it is the total
acceleration, internal plus external, that must be included in eq.\ 4 or  5. 
As a modification of gravity eq.\ 5 should read
$$\mu(|{\bf{g_i}}+{\bf{g_e}}|/a0)({\bf{g_i}}+{\bf{g_e}})
= {\bf {g_{N}}_i}+{\bf {g_N}_e}. \eqno(7) $$
where ${\bf g_e}$ is the modified external field, ${\bf g_i}$ is the
modified internal field, $\bf{ {g_N}_i}$ is the Newtonian
internal field and ${\bf {{g_N}_e}}$ is the Newtonian external
field.  
 
In other words 
the internal dynamics of
a system is influenced by the acceleration of any external
field beyond the usual tidal effect; if the external field 
has an acceleration greater than $a_0$, then the system 
is not in the regime of modified dynamics.  It is important to note 
that this was not a new assumption but
a natural consequence of the non-linearity of the formalism. The form
of the algorithm implies that
the underlying theory of MOND violates the 
strong statement of equivalence principle (although the
weaker version, the universality of free-fall, is of
course respected).  And, as is obvious, that theory could not
be General Relativity since GR embodies strong equivalence.

I show, in the appendix, a letter written in 1982 by Milgrom to John Bahcall,
at that time the head of the astrophysics group at the Institute for 
Advanced Study and an early critic of the idea.  Milgrom's letter 
makes a clear 
and insightful exposition of the philosophy and phenomenology
behind the idea and presages much of what had been written about
MOND since that time.  It is a remarkably up-to-date exposition.

\section{And then there were two}

Early on, in 1982, MOND received its first active supporter.
That was Jacob Bekenstein, a respected theoretical physicist
who was famous for proposing that black holes possessed the
property of entropy.  Together, Bekenstein and Milgrom 
worked out the first field theoretic version of MOND as
a modification of Newtonian gravity \cite{bm}.  They wrote down
a non-relativistic Lagrangian which possessed 
space-time translational
and rotational invariance (published in 1984).  Therefore the 
resulting theory respects the laws of conservation of 
energy and linear and angular momentum (this addressed issues
raised by several of referees of Milgrom's initial
papers).  The field equation
has the form of a non-linear modified Poisson equation, i.e.,
$$\nabla\cdot\Bigl[{\mu\Bigl({{|\nabla\phi|}\over{a_0}}\Bigr){\nabla\phi}}\Bigr]
= {4\pi G \rho} \eqno(8)$$
where $\mu$ has the asymptotic behavior described earlier.
Bekenstein and Milgrom demonstrated that the external field effect
was embodied by such a theory as was the center-of-mass motion
of a compound object in an external field.

Of course, the theory did not yield precisely the same
gravitational field as the simplified MOND algorithm
described by
eq.\ 5; only in cases of high symmetry, spherical or cylindrical,
did the Bekenstein-Milgrom field equation provide the 
the same acceleration field, $g$.  The non-linear equation was difficult to
solve, but, later on, it was demonstrated that for disk galaxies
the simple version provided a reasonable
approximation to the solutions of the modified Poisson equation
\cite{bradmil}.

Although non-relativistic, the Bekenstein-Milgrom theory 
pointed the way toward
a possible relativistic extension as a modified scalar-tensor
theory.  Here, gravity is mediated by two fields -- the usual
metric tensor of general relativity plus a scalar field with
an unconventional kinetic Lagrangian (also an aspect of what would
become known as k-essence). Given that $l_0$ is a length scale
on the order of the Hubble radius (with $a_0 = c^2/l_0$), 
and defining the dimensionless scalar field invariant as 
$$\chi = {l_0}^2{\phi_{,\alpha}\phi^{,\alpha}} \eqno(9)$$
the scalar action becomes 
$$S={1\over {2{l_0}^2}}\int{ F(\chi) d^4x}\eqno(10)$$
with $F(\chi) = \omega\chi$ in the limit of large $\chi$, the usual
scalar invariant where field gradients are large. Here $\omega$ is 
a large number that is equivalent to the Brans-Dicke parameter in the
Solar System.  When $\chi$ is small, the low acceleration limit, then
$F(\chi) = {2\over 3}\chi^{3/2}$.  Thus in
the weak field limit, the scalar field equation is equivalent
to eq.\ 8 above with $\mu(x) = dF/d\chi$.  This is combined with
the usual Poisson equation for the Newtonian potential to yield the 
complete theory in the weak field limit.

To preserve the universality of free fall (the weak statement of the
Equivalence Principle), the scalar field couples to matter as
a conformal factor multiplying the Einstein metric, 
as in Brans-Dicke theory.  Thus particles follow the geodesics
of this ``physical metric" that is conformally related to the
Einstein metric.

A possible problem was immediately identified by
by Bekenstein and Milgrom:  in the low gradient limit scalar waves propagate 
with a velocity in excess of that of light ($\sqrt{2} c$)
in a direction parallel to the field gradient.
Whether or not this is an actual problem leading to
causal anomalies has been a matter of debate, but there was
another, more serious, possible observational problem: 
it is evident that null-geodesics 
of conformally related 
metrics coincide.  This means the scalar field has no influence on
photons or other relativistic particles implying that
there would be no
enhanced deflection of photons by the scalar field -- MOND would
not provide extra bending of the path of light (photons would not
``see dark matter") in galaxies or clusters of galaxies for example.
Later observations would show that this was clearly not the 
case. 

Finally, the theory left the issue of cosmology hanging in the
air.  In the cosmological limit, where $\chi$ changes signs,
the form of $F(\chi)$ must be modified further.  Moreover, 
the constant $l_0$ does not follow naturally but must
be put in by hand.  Thus, the apparent significance of $a_0\approx cH_0$
is lost.

In spite of these issues, the suggestion of possible
relativistic extension of MOND was significant.  It demonstrated
that such an extension may be realized as a multi-field
theory of gravity, and this led, after 20 years, to the first
consistent covariant theory, TeVeS \cite{teves}.  

\section{The predictive power of MOND} 

I became an active supporter in 1985 when confronted with
the high quality 21 cm line rotation curve data then 
being provided by the Groningen group \cite{begem,valbetal} 
using the Westerbork 
radio interferometer (WSRT).  In principle, the MOND
algorithm permits the calculation of any rotation curve
using the observed distribution of baryonic matter (stars
and gas).  There is, of course, the uncertainty presented
by the unknown form of the interpolating function $\mu$, but
in practice, the results turn out to be rather insensitive
to the abruptness of the transition between modified and 
Newtonian dynamics.  Characterizing the function as 
$$\mu(x) = {x\over{(1+x^n)^{1/n}}} \eqno(11)$$ 
it has been found that $n=1$ or $n=2$ work about equally well,
although there is a difference in the implied mass-to-light
ratio of the stellar disk.  Moreover, a number of
low surface brightness gas-dominated galaxies are in the 
deep MOND regime and therefore independent of
the form of $\mu$ as well as the stellar mass-to-light ratio;
the distribution of baryonic matter is
given almost entirely by the directly observed 
gas surface density distribution.

\begin{figure}
\begin{center}
\includegraphics[height=12cm]{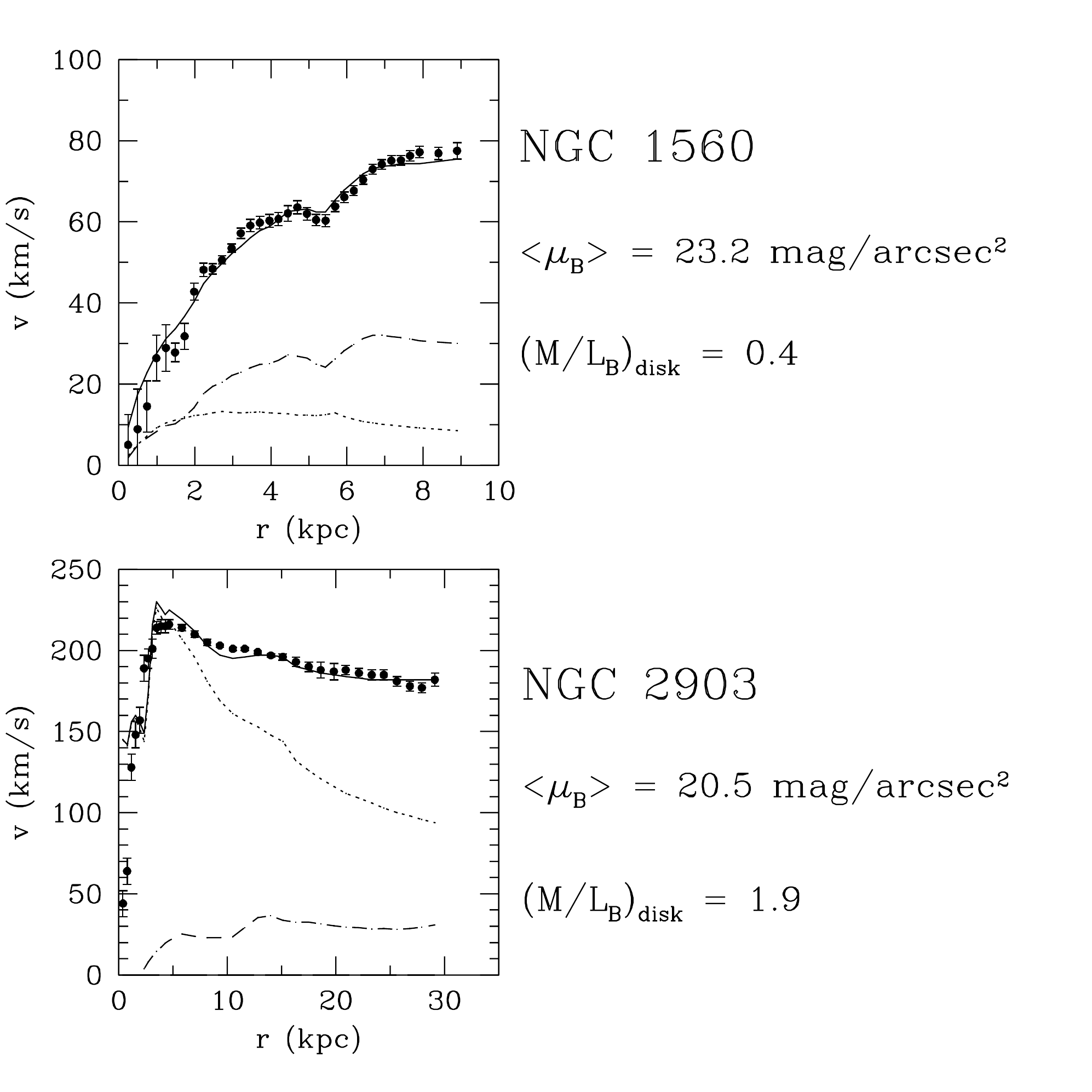}
\caption{The 21 cm line rotation curves of two spiral galaxies,
high and low surface brightness (HSB and LSB), determined with the
MOND algorithm from
the observed distribution of observable matter, stars and gas.
The points are the observed values,
the dashed curve is the Newtonian rotation curve of the gaseous
disk, the dotted curve is that of the stellar disk and the solid curve
is the predicted MOND curve.
The general form of the rotation curves was predicted 
by Milgrom in 1983:  the rotation velocity in the HSB declines
in a near Keplerian fashion to the asymptotic constant rotation
velocity, and the LSB the rotation curve gradually rises to
this asymptotic value \cite{bbs}.  
}
\end{center}
\end{figure}

The procedure is to estimate the distribution of
baryonic matter by surface photometry of the visible disk,
preferably in the near-infrared.  The gas is included
by taking the measured surface density multiplied by 
a factor (typically 1.3) to include the primordial 
helium.  The stars and gas are assumed to be distributed in
a thin disk, apart from those cases where there is direct
evidence for a spheroidal bulge.  The distribution of the
Newtonian gravitational acceleration 
is determined by solving the Poisson equation and then modified to
the MONDian (``true") gravitational acceleration by applying the
MOND algorithm (eq.\ 5).  The mass-to-light of the disk is
adjusted so that the fit is optimal.

The results of applying this procedure are shown in two cases
in Fig.\ 5 \cite{bbs}.
These are 21 cm line rotation curves of a LSB, low-surface-brightness, (NGC 1560)
and a HSB, high surface brightness galaxy, (NGC 2903), where the points
show the observations, the dotted and dashed curves are the
Newtonian rotation curves of the observed baryonic components (stars
and gas), and the solid curves are the MOND curves resulting from
the simple algorithm (with $a_0 = 1.2\times 10^{-10}$ m/s$^2$).
One should note that the rotation curves have the general forms
for LSB and HSB predicted by Milgrom in 1983:  in the LSB galaxy
the curve slowly rises to its asymptotic form and in the HSB
the curve falls in a near Keplerian fashion to the final value.
The required mass-to-light ratios (shown on the figure) are
quite reasonable for a gas dominated and a star dominated galaxy.
There are more than 100 such rotation curves now in the literature
\cite{sanmcg,fmcg} and in roughly 90\% of these, the agreement is of
comparable quality (not all rotation curves are expected
to be perfect because of uncertainties in the fundamental
parameters -- distance, inclination, warps, non-circular motions).

In principle there is one free parameter per galaxy: M/L ($a_0$ must be
the same for every galaxy).  For the uniform sample
of spiral galaxies in the Ursa Major cluster \cite{sanverh} 
the fitted mass-to-light
ratios agree well with those implied by population synthesis
modeling \cite{beldej} as is seen in Fig.\ 6.  Indeed with
improvements in stellar population synthesis modeling
and in gas dominated galaxies 
it has become possible to make zero-parameter ``fits",
i.e., true predictions.  

\begin{figure}
\begin{center}
\includegraphics[height=7cm]{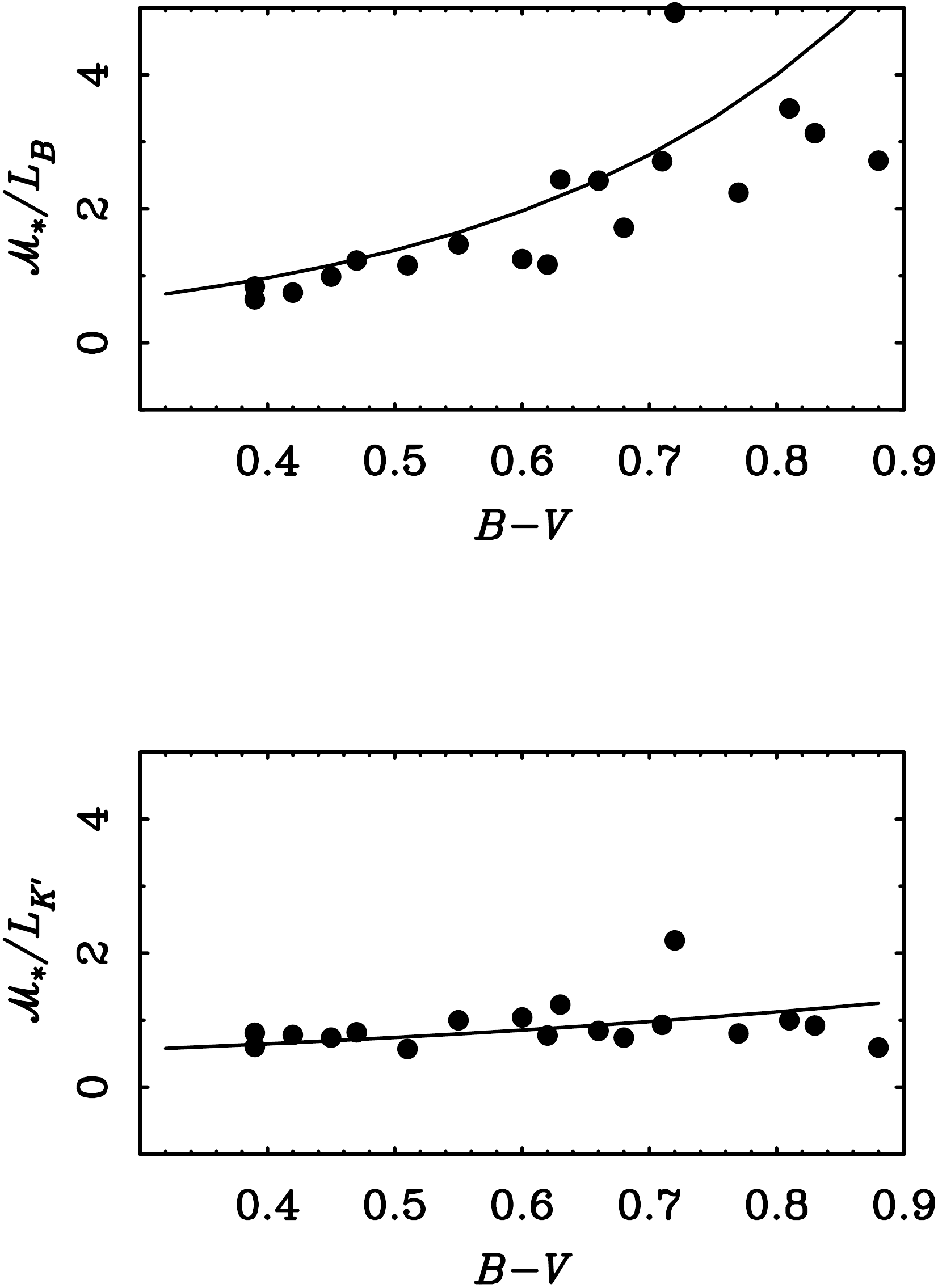}
\caption{The mass-to-light ratios for the Ursa Major
spirals inferred from the MOND fits
to rotation curves.  The top panel is in the B-band and
the lower panel is in the K'-band.  The curves are from
population synthesis models of Bell and de Jong []. }
\end{center}
\end{figure}

The remarkable aspect is that, in many cases, details in
the rotation curves are matched by the MOND rotation curves.
This explains an aspect of the observations noted by
Sancisi \cite{renlaw}:  Every feature in the observed distribution
of matter is matched by a corresponding feature in the rotation
curve, and vice versa, {\it even in the presence of a large
conventional discrepancy}.  If there is a cusp in the light 
distribution, there is a cusp indicated by the rotation curve as in
Fig.\ 7.  If there
is a fluctuation in the observed surface density distribution, as
for the dwarf galaxy in Fig.\ 5, there is a corresponding
fluctuation in the rotation curve.  In the context 
of dark matter
this would appear to be quite unnatural.  It would require a
coupling between the dark and visible matter that is not
implied in presumed nature of dissipationless dark matter. 
These results convinced me that
there must be something fundamentally correct about MOND.
Dark matter, as it is thought to be, could not possibly 
match this predictive success.

\begin{figure}
\begin{center}
\includegraphics[height=100mm]{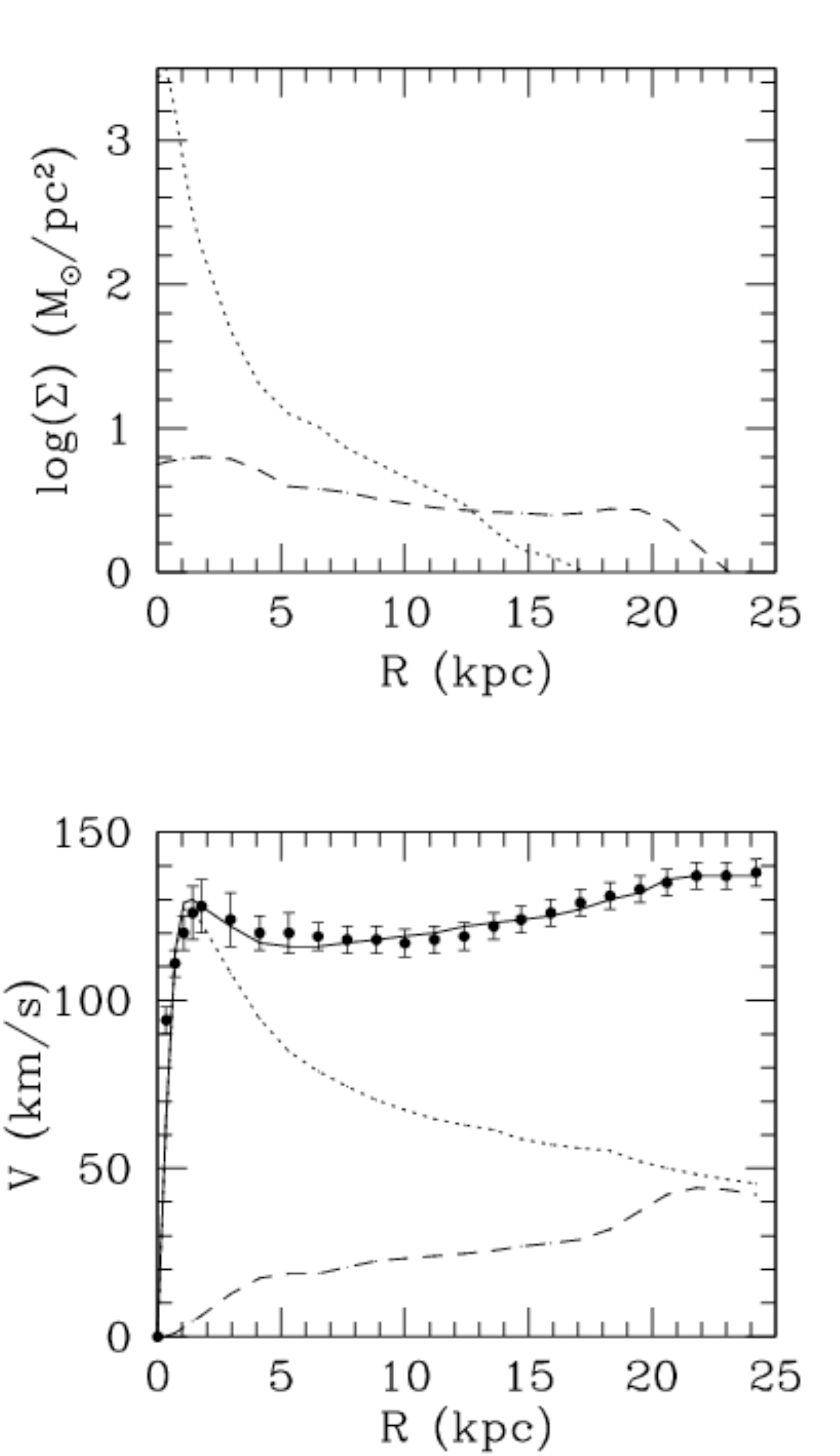}
\caption{Top:  the mass surface density distribution in stars and
gas (dotted and dashed curves) as a function of radius for the
dwarf galaxy, UGC 6406.  Bottom:
the corresponding Newtonian 
and the MOND rotation curves (dotted, dashed, solid).  
The points are the observed curve.  Note that there is
a cusp in the light distribution with a corresponding
spike in the central rotation velocity.  The rotation curve
slowly rises in the outer regions, not because of a halo, but,
in the context of MOND,
due to the increasing contribution of the neutral gas in this
low acceleration region.  From unpublished observations by
Zwaan, Bosma and van der Hulst (2005).}
\end{center}
\end{figure}

And recall that the same value of $a_0$ which provides these
fits, 1.0 to 1.2 in units of $10^{-10}$ m/s$^2$, also
follows as the normalization of the Tully-Fisher relationship.
This fact impressed early supporters such as Stacy
Mcgaugh.  The same Tully-Fisher law was present for galaxies
of low and high surface brightness \cite{mcgdb}.  When McGaugh plotted
the baryonic mass against asymptotic rotation velocity from the
measured rotation curves (not a global 21 cm line width), the
relation only improved (Fig.\ 4) \cite{mcgbtf}.

There were other subsequent observations that had been predicted by
MOND:  the ubiquity of a large discrepancy in low surface brightness
systems \cite{sanmcg}; the absence of a discrepancy in high surface brightness
systems such as luminous elliptical galaxies \cite{rom,milsan}
and globular clusters \cite{prymey};  
the presence of a preferred internal
acceleration in pressure supported systems ranging from
molecular clouds in the Galaxy to the giant clusters of galaxies --
an internal acceleration within a factor of thee of $a_0$ \cite{sanmcg};
the fact that systems ranging from globular clusters to
clusters of galaxies lie on the same $M\propto \sigma^4$ 
relationship \cite{sanfj};  the extension of Tully-Fisher (or Faber-Jackson)
to very large distance (i.e., small accelerations) indicated
by weak gravitational lensing \cite{millens}. 
And all of this was accomplished with a single value of $a_0$
having the cosmologically preferred value near $cH_0$.
Moreover, there have been no subsequent modifications of MOND.
It has not been necessary to tweak the theory 
to fit new and better data; in fact,
one has the impression that as the data improves, so does
the agreement with MOND.

\section{Criticisms and challenges} 

In his original paper on clusters \cite{milg83c} Milgrom, using
the MOND algorithm for pressure supported
systems, estimated 
the mass-to-light ratios of 15 clusters of
galaxies.  He noted that in about
half the cases the mass-to-light ratio 
remained unreasonably high ($\approx 20$
rescaling to $H_0=75$).  Milgrom pointed out several 
possible causes of systematic error including the substantial 
contribution of hot gas to the total mass
which had not generally been appreciated at that time.

In 1988, The and White \cite{tw88} in a more detailed
analysis of the Coma cluster, including the contribution
of the gas, noted that MOND did
not resolve the entire discrepancy;  that to
remove the need for unseen mass $a_0$ had to be more than
a factor of two times larger than that required to fit
galaxy rotation curves.  This can also be interpreted as
a statement that MOND requires more matter in the cluster
than can be directly detected.  Over the next 10 years
it became clear that the problem was more general --
that most clusters of galaxies, when analyzed with MOND,
contained more mass than was directed detected \cite{gerbetal}.  
In particular I found \cite{sand99} that there was typically a factor
of two to three remaining discrepancy in a large sample
of X-ray emitting clusters (see Fig.\ 8).

\begin{figure}
\begin{center}
\includegraphics[height=7cm]{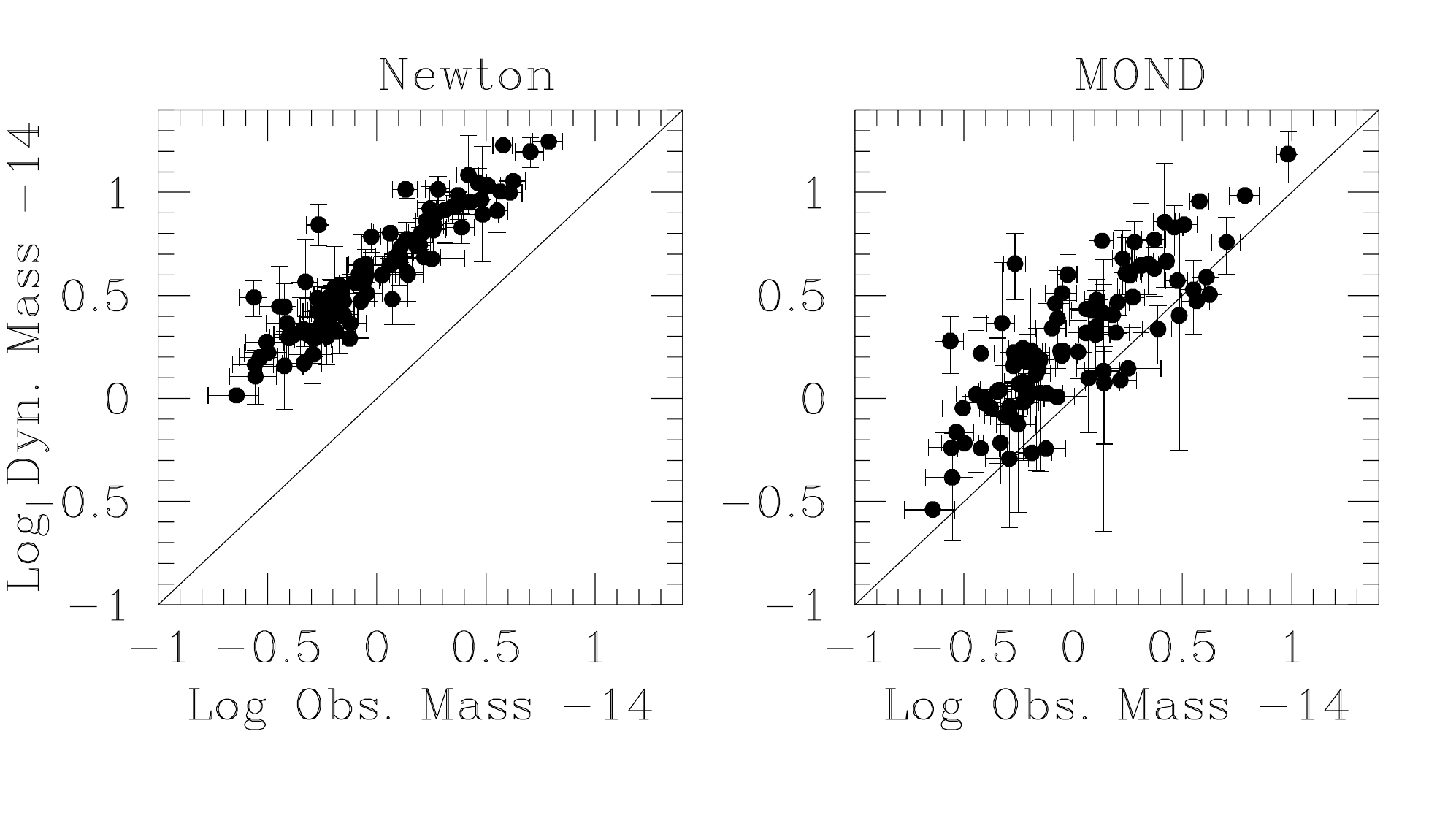}
\caption{The panel on the left is a log-log plot of the Newtonian
dynamical mass against the directly observed baryonic mass
(mostly in the form of hot gas).  The units are $10^{14}$ $M_\odot$.
The panel on the right is the same for the MOND\index{MOND, MOdified Newtonian Dynamics} dynamical mass.
Note that MOND\index{MOND, MOdified Newtonian Dynamics} reduces the discrepancy but does not remove it.}
\end{center}
\end{figure}

This has been trumpeted as a failure of MOND on scales
greater than that of galaxies.  These claims of doom 
culminated with multi-frequency observations of the famous "bullet"
cluster \cite{clowetal} in which the stellar components of
two colliding clusters have apparently
passed through one another leaving behind the collisional
X-ray emitting gas.  The putative dark matter, detected
by weak gravitational lensing of background galaxies, 
has also passed through and
coincides in position with the cluster galaxies.  

This observation
does present a challenge for MOND but no greater than the
older problem of the remaining virial discrepancy.
Formally, the observed remaining discrepancy is not 
a falsification of MOND; it would
be if MOND predicted less matter than is actually seen.
In fact, one could view this a bold prediction of MOND:
more mass will be detected in clusters.

As pointed our by Milgrom \cite{mcoldcls} there
are more than enough undetected baryons in the Universe
to make up the difference.  If this constitutes the
undetected cluster mass, then bullet cluster result implies
that the baryons must be in some, effectively, dissipationless
form -- small, compact, cold clouds, for example.
Another possibility is that of massive neutrinos, either
the standard three \cite{sandneut1} or a new sterile neutrino \cite{angus}.  
If the three standard
neutrinos have, for example, masses on the order of
one to two electron volts, then, due to phase space 
constraints, they can accumulate on the
scale of clusters but not that of galaxies. For standard
neutrinos this issue will be settled soon because direct
beta-decay measurements of the masses (or relevant
limits) are currently underway \cite{wolfk}.

From the beginning MOND has been criticized for its 
absence of a relativistic extension.  As a non-relativistic
theory, MOND (or the Bekenstein-Milgrom theory), 
makes no predictions with
respect to relativistic phenomena, such as
gravitational lensing or an alternative cosmology.  
Again this shortcoming is an incompleteness, but it is not a
falsification.

Felten, in his paper on dynamical problems with the original
MOND formulation \cite{felt}, first described the potential problems
with a MONDian cosmology, at least with the quasi-Newtonian 
treatment.  He pointed out that the expansion in a region
dominated by MOND cannot be uniform; separations cannot
be expressed in terms of a universal scale factor.  
This means that
any such region will eventually re-collapse regardless of
its initial expansion velocity or density (Felten
did note a positive aspect of this property: structure
formation is hierarchical from bottom up and that
the size of region now separating from the Hubble flow
would be on the order of 30 Mpc, comparable to the
scale of large scale structure).  Felten's argument
does not represent a failure of MOND, but more likely the
absence of a Birkoff theorem.  This theorem, arising 
in the context of general relativity,
justifies the application of Newtonian dynamics to a
uniform spherical expanding region in deriving the
Friedmann equations.  In the deeper theory of MOND 
a missing Birkoff theorem
would seem to be consistent with the external field effect. 

Applying the 
MOND non-relativistic formalism only to 
density fluctuations does permit a standard cosmology but with
a much enhanced growth rate for the inhomogeneities \cite{sandfluct,
nusser}.  
This all remains tentative, however, in the absence 
of a relativistic theory.  Is, for example, the acceleration
parameter constant or does it vary with cosmological time?

Thus, the essential challenge for MOND has been and remains the
absence of a more basic theoretical underpinning of the
concept -- a relativistic theory.  There are various
candidate theories, to be discussed further in this
compendium, but I will summarize several historical
developments (see also the up-to-date review in \cite{fmcg}).  

Most of the trial theories, following the original Bekenstein-Milgrom
relativistic toy theory, fall in the category of multi-field
theories.  Here the gravity force is mediated by two fields --
the usual tensor field of general relativity and a second field,
most often, a scalar field.  This is illustrated in Fig.\ 9 where the
force from the usual GR-Newtonian field is compared with that
of a non-standard scalar field.  The two fields become
equal in strength at the MOND radius, $r_0 = \sqrt{GM/a_0}$., but
the acceleration due to the scalar field must return to
$1/r^2$ attraction at higher accelerations in order to
satisfy Solar System constraints.

\begin{figure}
\begin{center}
\includegraphics[height=7cm]{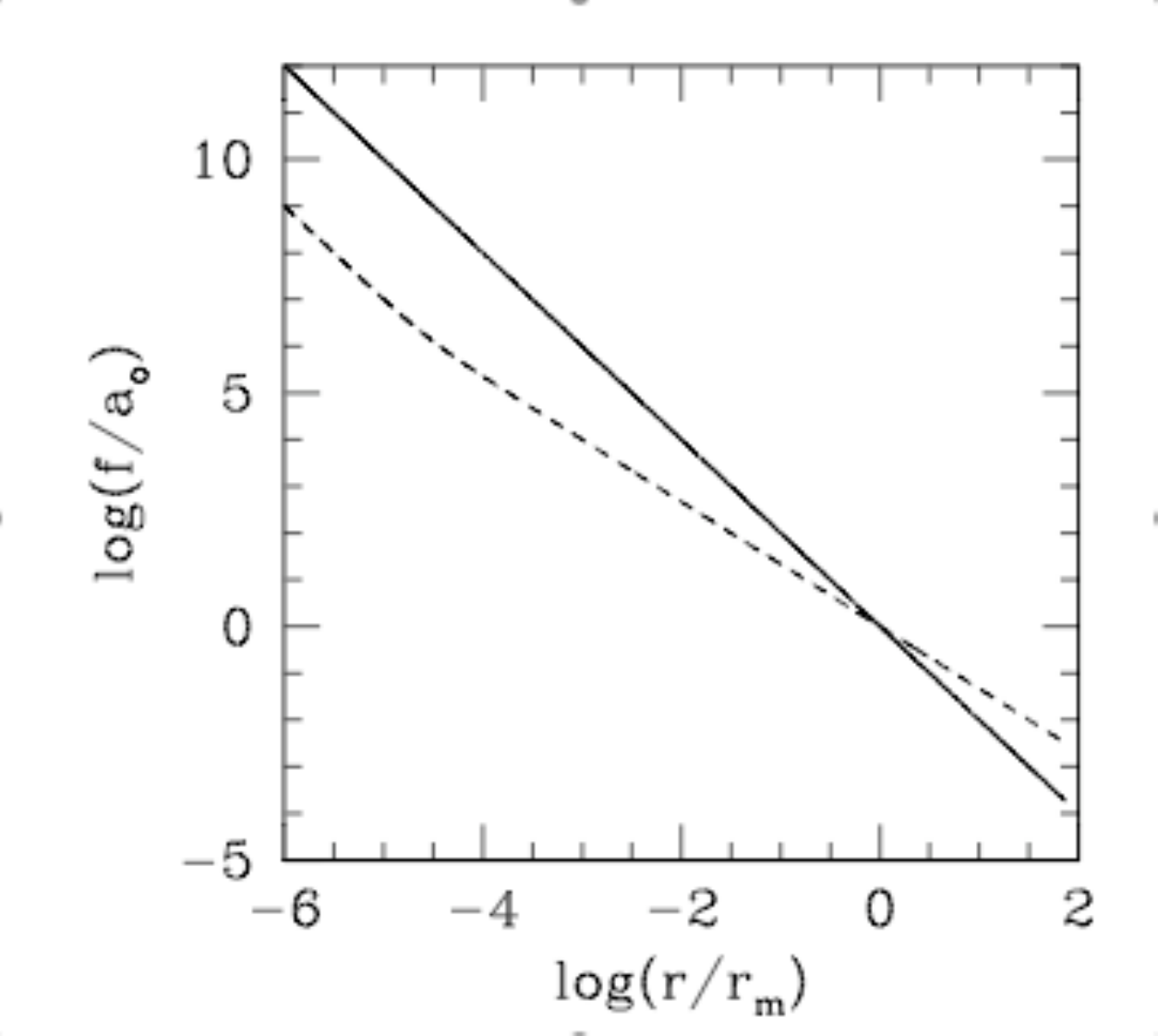}
\caption{A log-log plot of the run force per unit mass due to
the traditional Newton-GR field (solid curve) and due to a
non-standard scalar field that produces MOND phenomenology (dashed curve).
The two forces become equal at the MOND radius $r_m = \sqrt{GM/a_0}$.
At high accelerations the scalar force must also develop a $1/r^2$
dependence to satisfy Solar System constraints on deviations from
inverse square attraction.}
\end{center}
\end{figure}

In 1994 Bekenstein and I \cite{bs} addressed the
problem posed by enhanced gravitational lensing about
clusters of galaxies, as was evident in the observations
by that time \cite{fm}.  As mentioned above, a traditional
scalar-tensor theory fails to produce lensing beyond 
that of general relativity (without dark matter) due
to the conformal relation between the physical and Einstein
metrics.  A conformal transformation takes the geometry
of a particular space-time and expands or contracts it 
isotropically in a space-time dependent way.  Bekenstein
(1993) had earlier 
realized that a {\it disformal} relation between
the metrics could be a way around the problem of lensing \cite{bek93}.
A disformal transformation picks out one direction in a 
preferred frame as special 
for additional contraction or expansion, and null 
geodesics of the original and transformed metric
do not coincide.  In 1997 I proposed 
a specific theory that realized such a transformation \cite{strat} -- a
theory with a non-dynamical vector field pointing in
the direction of cosmological time, thus breaking the Lorentz
Invariance of gravitational phenomena as well as
time-reversal invariance (this was actually a modified form
of historical alternatives to General Relativity
known as ``stratified theories" \cite{strato}).  I demonstrated that,
with a particular form for the disformal transformation,
one could provide a scalar-tensor theory for MOND that yielded
the same relation between deflection of photons and
the total weak field force (including
the MOND force) as in General Relativity; that is to say,
with respect to lensing the theory was equivalent to
GR with dark matter.  

The problem was that the non-dynamical
vector field -- a field that acts upon 
matter and other fields
but is not acted upon -- violates the spirit and letter of
covariance with bad consequences for conservation principles. 
This deficiency was corrected several years later by
Bekenstein in his fully covariant theory
TeVeS (tensor-vector-scalar theory) \cite{teves}.
As in the stratified theory, TeVeS 
makes use of two fields in addition to
the usual metric tensor of General Relativity -- a
scalar field to provide the MOND force and a
vector field to provide a disformal relationship between
the Einstein and physical metrics.  But the vector field
is fully dynamical and his its own field equation and source.
Although the theory is not without problems, phenomenological
and conceptual, it demonstrates that a fully covariant 
theory of gravity leading to MOND in the weak field
limit can be constructed. [One such conceptual problem
is that the theory does not reduce to General Relativity
as the acceleration constant, $a_0$, approaches zero.]

There have been a number of successors to TeVeS which
attempt to address some of the perceived problems.
For example, I introduced a bi-scalar vector-tensor theory
to provide a cosmological interpretation of
$a_0$ along with cosmological dark matter in the form of
long wavelength bosons that do not cluster on the scale 
of galaxies \cite{biscal}. Einstein-Aether theories (vector-tensor
theories) \cite{jm} can be adapted as a relativistic theory
of MOND; here a function of the
vector field invariant is added to that of the tensor
field to promote MOND phenomenology \cite{zlosetal}. One advantage
of EA theories is that a cosmological term naturally
appears with a value on the order of ${a_0}^2$. Bimetric
theory has also been suggested \cite{bimond}; here the 
difference of the two Levi-Civita connections (also a tensor)
appears directly as an acceleration, normalized by
$a_0$, and is added to the action that includes the Ricci
scalars of both metrics.  It can be shown that this 
theory also produces a cosmological
term of order ${a_0}^2$ and reduces to 
Einstein-Aether theory when the two metrics are disformally 
related via
a unit vector.  I mention also non-local single metric
theories \cite{souswood} and dipolar dark matter theories 
\cite{blanchlt}.

In an entirely different vein there is also work building
upon the interpretation by Verlinde \cite{verlinde} of the Newtonian
attraction as an entropic force due to microscopic degrees of
freedom on a holographic screen.  When a maximum
scale corresponding to the de Sitter horizon is introduced 
for the screen, 
the gravitational force is modified in a form equivalent
to MOND with $a_0$ being identified with the 
Unruh temperature of the horizon (see \cite{klinkkop} and references
therein).  
This is an interesting suggestion but has not
yet been fully developed.

It is clear that there is at present no shortage of theories, but
at most one of these is correct.  Until this is resolved
by confrontation with cosmological observations, MOND
cannot progress further as an alternative to the 
present paradigm on the largest scale.  

\section{Conclusions:  Crazy ideas and MOND}

As a field, astronomy is replete with unconventional
characters who propose and become obsessed with bizarre
theories (often after a distinguished career as a
conventional scientist).  One example is provided by
Victor Ambartsumian, a well-known Armenian astrophysicist
who, in the mid-1950s, began speculating about the role
of galactic nuclei in the formation and evolution of
galaxies \cite{ambart}.  He suggested that small galaxies are born
complete from larger galaxies -- emerging from the
nucleus like Athene springing
full grown from the head of Zeus.  Clusters of galaxies
are not gravitationally bound systems but recently born
galaxies expanding away from the large parent galaxy in
the center; given a few billion years, the whole configuration
will dissipate into intergalactic space -- thus solving 
the mass discrepancy problem in clusters.  Although there 
were international meetings in which the idea was 
discussed, most astronomers did not take Ambartsumian's 
proposal seriously.  It was just too ``crazy".
[It interesting to note, however, that Ambartsumian's suggestion
that the nuclei of galaxies have a dominant effect 
on the evolution and morphology of galaxies has re-emerged
recently in a different guise with a different vocabulary --
``feedback" in which activity of the central massive black hole 
limits the growth of the visible object.] 

Every astronomer is aware of other examples, particularly
in cosmology, and
given the abundance such
bizarre proposals, it is an effect that we should be wary of.
So in the context of the present discussion the question arises: 
does MOND fall into the category
of crazy ideas?  From the strong emotions that the mere mention
of the word evokes at times, it would certainly seem as though 
some distinguished scientists think so.  But here, I will argue that,
while MOND is unconventional and 
inconsistent with the current cosmological paradigm, it is by no means
in the category of crazy ideas.  And we should recall that 
many constructs of modern physics, such as quarks, were at
an early stage considered crazy and condemned quite viciously
by renown scientists. [George Zweig, one of the creators of
the concept of quarks wrote: ``The reaction of the theoretical
physics community was generally not benign ....
When the physics department of a leading university was
considering an appointment for me, their senior theorist,
one of the most respected spokesmen for all of theoretical
physics, blocked the appointment at a faculty meeting by 
passionately arguing that the model was the work of a
charlatan." (quoted by Harold Frisch \cite{frisch}).]

Robert Ehrlich, in his book {\it Nine Crazy Ideas in Science} \cite{robehr}
has listed several questions to be answered in order to tell
if a crazy idea just might be true?
I paraphrase these here as a list of criteria for an idea to be
taken seriously.

1.  {\it The new hypothesis should make some contact with familiar
physics.  Well established physical principles
or some reasonable extension of those principles should be
respected.}  It is not, for example, easy to make such
a connection for the idea of non-cosmological redshifts of
quasars.  But we have seen that MOND makes plausible 
modifications of current physical law in a regime
where this law has never been tested and, certainly in its
Lagrangian form, does embody cherished principles.  

2.  {\it The proposer of the idea should be a knowledgeable
and respected scientist, although he/she may come from
outside that particular field.}  The world is full of
gifted amateurs who imagine that they have found the key
to a particular problem (or the theory of everything),
but almost never are they near the truth.  Again MOND
and its creator, along with the several respectable scientist
who have contributed to its development, meet this condition.

3.  {\it The proposer should not be overly attached to the idea.}
I would not rank this condition so highly because, 
actually, it is difficult not to become
attached to an idea that one believes is correct and 
not to feel somewhat defensive when the idea is 
ignored by a majority of the relevant community.  More relevant
is for the proposer not to be dismissive of observations or
data that does not support the idea.  I believe that the
supporters of MOND in general have not tried to sweep anything
under the rug, although perhaps they have been overly 
dismissive of cosmological observations that are now quite
precise.

4.  {\it Statistics should be applied in an honest way.}  
The use of {\it a posteriori} statistics with
respect to non-cosmological redshifts is an example of
misapplied statistical arguments, but this is not so
relevant to the problem of the mass discrepancy in
astronomical objects.  Of course, scaling relations such
as Tully-Fisher are statistical in nature, but the
most stringent and relevant selection criteria have been
applied to data such as that plotted in Figure 4; for
example it is the asymptotically constant value of the
rotation velocity, beyond the visible disk, that is plotted
rather than the width of a global line profile.

5.  {\it The proposer should have no agenda going beyond the
science of the issue.}  This is more relevant to fields
with political or economic impact, such as the reality 
(or not) of global warming;  it does not apply
here.

6.  {\it The theory should not have many free parameters.}
MOND has exactly one new fixed free parameter -- $a_0$
the critical acceleration which has, coincidentally, a 
cosmologically interesting value.

7.  {\it The idea should be backed up by references to
other independent work.}  Here again, the tests of MOND are 
generally based on
data -- kinematic and photometric -- taken by
independent, objective observers with no personal
stake in the theory.  In fact, they are most often
negative about the idea.

8.  {\it The idea should not try to explain too much or too little.}
Most of us have received emails from amateurs who attempt
to explain dark matter, quasars, the Big Bang, solar
neutrinos and the frequency of earthquakes with one
grand theory.  Such theories cannot actually calculate 
anything or make definite predictions.  On the other hand,
if the theory is
too narrowly focussed on, for example dwarf spheroidal galaxies,
then it lacks the generality appropriate to a physical 
theory but could more properly be described as a model.
MOND, I would say, strikes the right balance in addressing the
mass discrepancy in astronomical systems, although its 
implications would certainly be more general.

9.  {\it The supporters should be open about their data and
methods.}  With respect to MOND, nothing is hidden.
The data, photometric and kinematic, are published by
others 
and generally available.  The methods are simple, clear 
and reproducible by anyone.

10.  {\it The theory should provide the simplest explanation
of the phenomena.}  There should not be too many 
epicycles necessary to save the phenomena.  MOND comes
out very well indeed under this criterion.  It has never been
modified with new constructs or parameters in order to
explain those phenomena which it addresses, even though
observations have become more precise and the data have 
improved considerably.  The dark matter 
hypothesis, on the other hand, has required numerous
tweaks and adjustments to explain the well-known
discrepancies -- the cusp-core problem, the missing
satellites, the galaxy mass function.  

To this list I would add two more points:  
{\it First of all, it is
significant if the idea is falsifiable.}  Are there 
observations that can disprove the theory in question?
For MOND this is certainly the case; if, for example,
the algorithm predicted less matter in clusters than
is actually observed, this would be a certain 
falsification.  For galaxy rotation curves, if a number of these required
mass-to-light ratios that were negative, then this would be 
a falsification.  {\it Secondly, it is significant if
the number of proponents increases with time, especially
if the new converts are younger scientists.}  It does
not bode well for a theory if its principal supporters 
comprise a declining group of embittered old men.  This is not
the case for MOND, where the number of advocates, especially
younger scientists,
has doubled or tripled over the past 10 years.

So MOND meets most, if not all, of Ehrlich's (and my) criteria for
an idea to be taken seriously.  Why then has it languished
for more than 30 years outside of the mainstream?  After
all, quarks became an accepted concept within a few years
of being proposed.  

This is not due to some grand conspiracy.  There was,
however, from the first appearance of MOND an alternative
and dominant paradigm that claims to account for
the same phenomena.  There are strong
social factors that maintain support of the
prevailing paradigm: an overriding tendency for scientists
to work within the established framework and to select 
data that reinforce rather than challenge it (an
effect that is supported by
competition for academic positions and
grants);  and significantly, in this case, there is  
the general reluctance of most astronomers
to tamper with historically established
laws of nature -- in part a reaction against the plethora
of crazy ideas in astronomy.

But beyond
this there is a tendency for astronomers to regard 
cosmology as the queen of sciences -- a
reductionist undercurrent that gives cosmology
priority over mere galaxy phenomenology.  
Data such as the pattern of anisotropies in the CMB
are very well-fit by the standard cosmological model, albeit
with a somewhat strange combination of six parameters.  
Given the precision of the fit, then the theory must be
correct, even in its implications for galaxies.  This
makes most cosmologists dismissive of galaxy phenomenology
and its wealth of regularities.  These are problems that
will be understood in the context of more detailed
computations of the baryonic processes in galaxy
formation and evolution -- problems for the future.

But viewed strictly as a epistemological issue --
without prioritizing classes of data -- the
CMB anisotropies are defined by a single curve, the angular
power spectrum, which
can be fit by a six parameter model.  But
there are at least 100 well observed galaxy rotation
curves which, when
combined with population synthesis models of the
stellar populations of
galaxies and thus color related mass-to-light ratios, 
can be fit by a theory having one fixed universal
parameter.

For MOND, the absence of a more basic theoretical
underpinning of the idea remains the essential weakness.
This means that cosmological calculations and predictions
must be deferred.
Ideally, one would like to have a theoretical determination
of the function $\mu$ that interpolates between the
Newtonian and MONDian regime -- at least for the phenomenon
of rotation curves.  The near coincidence of
the acceleration parameter $a_0$ with $cH_0$ must be
significant but there is not yet a convincing explanation.
Does $a_0$ vary with cosmic time, as $cH_0$?  Or is it
an aspect of a fixed cosmological constant $\Lambda$?
(In some theories, such as the Einstein-Aether version,
$\Lambda$ would appear to have a scale on the order
of ${a_0}^2$.)  Is MOND more properly a modification of
gravity or of inertia?  In general relativity these are two sides of
the same coin, but is that equivalence broken at low
accelerations? 

I have argued that the success of MOND on the scale of
galaxies strongly challenges the concept of
a dissipationless dark fluid that clusters on these
scales.  But are there other problems with the
standard cosmology -- problems that manifest themselves
on cosmological scales and times?  It is certainly
impressive that, more or less, the same set of 
parameters emerge from different observations:  
the CMB anisotropies, the recent expansion history as
traced by distant supernovae, the power spectrum of
matter fluctuations.  But rather than the precision with
which the parameters of a model Universe are determined, 
it is the peculiar
composition and the remarkable coincidences embodied
by the concordance model that call for deeper insight.
These motivations for questioning a paradigm are not
unprecedented;  such worries led to the inflationary
paradigm that
has had profound impact on cosmological thinking
over the past 30 years.

A more practical difficulty is the absence of an independent
detection of dark matter particles.  In the context
of $\Lambda CDM$ they should be abundant locally --
they cluster in the Milky Way.  Where are they?
Because the properties of hypothetical particles are limited
only by the human imagination, the concept of dark matter
is fundamentally not falsifiable, but at some point
continued non-detection must become a worry.  This is
an issue that is at least as problematic for dark matter as the
absence of a cosmology is for MOND.  Without independent
detection the concept of cold
dark matter clustering on galaxy scales remains
hypothetical, and the standard cosmological paradigm, 
upon which it is based, is a pipe dream.

All of this illustrates the dangers to the creative process
in science presented by dogma too widely and too deeply accepted.
In the context of the standard paradigm most work on galaxies -- 
for example semi-analytic galaxy
formation models -- is built around patching up the standard
model to make it work rather than challenging it.   
Indeed, it is difficult
to imagine that a fundamental challenge could emerge from
techniques characterized by numerous adjustable parameters
or effects newly added as needed.  
MOND presents a different, and more
traditional, sort of science in which definite predictions are
made and verified -- or not.  This is why the idea remains
and gains support.

I thank Moti Milgrom and Stacy McGaugh for helpful comments on
the manuscript.

\section*{Appendix}

\vspace{0.5cm}

Below I reprint, with permission of Moti Milgrom, his
response to a letter from John Bahcall concerning 
Milgrom's three preprints circulated in 1982.  Bahcall
had doubts about the timeliness of these papers because
he felt that there was no crisis with the hidden matter hypothesis.
This is one of several issues that Milgrom discusses in this
very insightful letter.  His points are philosophical and
phenomenological and presage comments by myself and
others made some years later on.  I make no further comment
upon this remarkable document.

\vspace{0.5cm}

***********************************************************************

\small{
\noindent April 4, 1982

\vspace{.5cm}

\noindent Prof. John Bahcall

\noindent Institute for Advanced Study

\noindent Princeton N.J. 08540

\noindent U.S.A.

\vspace{0.5cm}

\noindent Dear John,

\vspace{0.3cm}

Many thanks for your frank letter of Febr. 26.  I was
disappointed to read that you have not read at least
the more detailed paper on galaxies.  I do not
think it is possible to appreciate the work after
reading only the shorter paper.

I did not write the shorter paper as an independent
piece.  In it, I meant to discuss mostly matters of
principle and mention only briefly the results of the
other papers, in the hope that it will serve as an
appetizer.  Obviously, for you it did not.

As I already wrote to you, the flatness of galaxy
rotation curves was the first hint for me, some
3 years ago, that hidden mass may not be a satisfying
explanation.  Since then, during my stay in Princeton
and after that, I have studied the data very
carefully.  If you could see through the dust (that
no doubt exists), as I think I can, I think you would
be much more positive.  If one prefers to wait until
much of the dust settles down, one will see what there 
is to see with everybody else.

I think I am very careful with the data.  I have not invented any of the regularities which I take to exist in the data
and I think I have a good eye for things in the data which call for an explanation.  Knowing what I know about the data,
I feel very strongly that a new explanation is needed.

If I understood correctly, you think that the time has not
yet come to consider alternatives to the Newtonian dynamics
as you see no crisis emerging from the data as you now
understand it.  At the time of Copernicus you could argue
that the time has not come to consider alternative to the
Ptolemeyan system as it explained the data very well if you  were willing to take enough epicycles.  Whether the time has come or not is very much a matter of taste.  I am sure you will find many that are deeply dissatisfied with the hidden mass, as I personally am.

It should be realized that to say that the dynamics in galaxies and systems of galaxies are explained by hidden mass, which cannot be detected otherwise, is really saying practically nothing.  I have not proved this as a theorem, but practically any set of dynamic measurements is describable in terms of some mass distribution, in the framework of Newtonian dynamics (as long as you need more mass than you see).  What I mean is that you will never be able to rule out the HM ("hidden mass") hypothesis from
dynamical measurements.  You may be able to rule out certain forms of matter because they may have this or that extra effect, but you can always find shelter in unknown type of mass, etc.

What this also means is that it will be extremely difficult for the HM hypothesis in its general form to make any predictions and in fact, to the best of my knowledge, it does not make any testable predictions (other than those which may have to do with the nature of its constituents).

In addition, it can be said that none of the observed properties of galaxies or systems of galaxies are explained by the HM, in the sense that none follows in any way from this hypothesis and perhaps other known properties of galaxies (not the flat rotation curves, nor Tully-Fisher, nor 
the Oort discrepancy or the universal surface brightness).

What would make me seriously consider abandoning the HMH would be an alternative which
a. does not clearly conflict with any known experiment or observation.
b. explains many of the properties which the HMH does not.  Namely 
that these results follow from the alternative theory and
c. makes well defined predictions which can be tested.

Now, I think that my scheme has these properties.  I do not expect one to accept if after reading my papers (certainly not before reading them), but I think that I can expect that it will be put to further consideration and test.

Whether the time has come to consider alternatives is not a good question to ask in this situation.  I am suggesting an alternative that is a result of careful thinking and a lot of work.  The question to ask is whether it stands the test or not.

You say in your letter that one could think of many modifications of the present physics which will explain the data.  I do not think that this is so.  Again I thought about this a lot and I believe the data is more than enough to narrow down the possibilities very strongly.  It is true that the particular form of the modification I use is one of a few possible ones that I can think of, but they all share a few features which are the basis of my suggestion and from which most of the results follow.  They are: 1. a breakdown of Newtonian dynamics (second law and/or gravity). 2. the quantity which matters is the acceleration (kinetic or gravitational). 3. there exists an acceleration constant which determines the border between the Newtonian and 
non-Newtonian regime. 4. in the extreme non-Newtonian regime the square of the acceleration is proportional to the usual gravitational force.  From these follow the flat rotation curves, Tully-Fisher, preferred value of the surface brightness, a discrepancy in the z-dynamics, etc.  Details such as the exact shape of the rotation curves, the exact value of the Oort discrepancy, etc. depend on details of the modification.

You also complain in you letter that I do not really give a theory.  You could in the same vein complain to Bohr for not suggesting the Schroedinger equation together with his model for the atom (to remind you it took 13 years and many people much better than myself to get from one to the other).  I could give many more examples although I do not really think that they mean much, as one can always bring counter examples.  I think that each case should be considered on its own.

The important things are that I do make predictions and I do find relations between various properties of galaxies, and that known properties of galaxies follow unambiguously from my proposed scheme.  Take for example $a_0$.  It clearly has life of its own independent of my proposal.  It appears directly from the observations as 1. a universal average acceleration in spirals and ellipticals (as reflected in the preferred value of the surface brightness).  2. It appears as the proportionality factor in the Tully-Fisher relation for discs and ellipticals.  Looking at things my way I noticed that this parameter of galaxies is also within the uncertainties
the typical cosmic acceleration $cH_0$.  Maybe this is a coincidence but surely it is an interesting result.

To make my point clearer perhaps, suppose that my predictions are born out, namely using an expression similar to mine
with some $a_0$, rotation curves of individual galaxies are obtained correctly, the discrepancy in the z-dynamics is described correctly, deduced masses of systems of galaxies are equal to the observed ones, etc.  One may still want to maintain the HM, as it too produces the observations if the HM is distributed properly.  He will, however, have to agree that I found a {\it single formula} with one parameter which describes the distribution of the hidden mass in all galaxies and systems of galaxies (including the Oort-type mass) in terms of the observed mass distribution.  In addition the parameter $a_0$ happens to equal $cH_0$.  I would consider such a view an absurd one, and would much prefer to say that the formula really describes a modification of the physics and that there is really not much hidden mass.

I hope to see you in May and discuss matters in detail.

\vspace{0.3cm}

\noindent Best regards,

\vspace{0.5cm}

\noindent M. Milgrom
}

*************************************************************************
\end{document}